# Measuring the surface stress polar dependence

J.J.Métois, A.Saùl, P.Müller

Centre de Recherche sur la Matière Condensée et les Nanosciences[1]
Campus de Luminy, case 913, F-13288 Marseille Cedex 9, France

**Abstract:**

*While measurements of the polar dependence of the surface free energy are easily available, measurements of the whole polar dependence of the surface stress of a crystal do not exist. In this paper is presented a new procedure that allows, for the first time, the experimental determination of the surface stress polar dependence of a crystal. For this purpose (1) electromigration is used to control the kinetic facetting of surface orientations that belong to the equilibrium shape of the crystal (2) for each destabilised surface, the period of facetting as well as the crystallographic angle of the appearing facets are measured by AFM. The so-obtained data lead to a set of equations whose mathematical solution, compatible with physical constraints, give access to the surface stress polar dependence of the whole crystal and thus to a better understanding of surface stress properties.*

Two distinct macroscopic quantities are needed to characterise the thermodynamics properties of a crystalline surface n characterised by its normal direction $\hat{n}$: the surface free energy per unit area $\gamma^n$ which measures the cost of creating a surface area at constant deformation and the surface stress $[s]^n$, which measures the energy cost for deforming a surface at constant number of surface atoms. While $\gamma^n$ is a positive scalar, the surface stress which depends upon the direction of stretching the surface, is a second rank tensor. If the surface tends to shrink (resp: expand) in one direction (with respect to the bulk) the corresponding surface stress component $s_{ij}^n$ is positive (resp: negative) and is said to be tensile (resp: compressive). Since, in vacuum, a fully relaxed surface has no normal stress, the surface stress can be simply considered as a two-dimensional tensor which, when diagonalised, is simply characterised by two orthogonal components. Because of the anisotropic nature of the crystalline state, surface stress as well as surface free energy depends upon the orientation of the crystalline surface. However while measurements of the polar plot of the surface energy (that means its dependence with the surface orientation $\hat{n}$) can be easily obtained from a detailed study of the equilibrium shape of 3D crystals [1,2], measurements of the polar plot of the surface stress of a crystal are still lacking [3,4]. This is the most puzzling that during the last decade there has been an increasing interest about the influence of surface stress on many physical properties such as surface relaxation or reconstruction [5-8], surface segregation [9], surface adsorption [10], nano-sensors properties [11] and self assembling [12,13]. Moreover, surface stress is very often the main driving force for bottom-up nanostructuration [14]. *The new procedure we propose enables for the first time the experimental determination of the surface stress polar plot of a crystal.*

---

[1]Associé aux Universités Aix-Marseille II et III.

Since the equilibrium shape of a free crystal is the one which minimises its surface free energy, it is intuitive that the geometrical description of the equilibrium shape (ES) of a crystal is an image of its surface free energy anisotropy [15]. Thus, the polar plot of the surface free energy of all the surface orientations that belong to the equilibrium shape can be obtained from a simple inverse geometrical construction (In Fig. 1 is given the Silicon ES and its corresponding $\gamma$-plot [2]). Missing orientations on the equilibrium shape (ES) correspond to unstable surfaces. If such an unstable surface is macroscopically prepared (by slicing a crystal for instance), then annealed (to equilibrate it) it breaks up into facets of the two neighbouring orientations present on the equilibrium shape of the crystal [16]. The period of this spontaneous facetting is connected to the surface stress difference between the two stable orientations [12,13] which thus can be extracted from a simple wavelength measurement [17]. However, this procedure cannot be used for the crystallographic orientations that belong to the equilibrium shape which, since stable, do not facet upon annealing and thus, give no information on surface stress. Nevertheless, we show that this problem can be get round each time it is possible to use an external constraint to destabilise these stable surfaces. It is in particular the case of the stable vicinal faces of W [18], Ta, Mo, Pt, Fe, Ni [19] and Si [20] which can be destabilised by adatoms migration leading to step bunching instability then to kinetics facetting [20-24]. *In this work, such electromigration driving force is used to destabilise some well chosen vicinal facets in order to obtain the surface stress anisotropy of all the directions that belong to the equilibrium shape.* Because of its technological importance we have chosen silicon for which a good knowledge of the surface stress anisotropy should be welcomed !

For our purpose, we have selected the (118), (223) (443) and (105) stable vicinal faces (whose orientations were controlled by Laue diffraction) of a Silicon crystal (see Fig. 2) and heated them (in ultra high vacuum conditions) by Joule effect (AC or DC current). When using AC current, all the vicinal surfaces remain flat, while using DC current in the direction of ascending steps[2] in the range 1200-1400K the surfaces break up into a hill and valley structure (see Fig. 3) [20-22][3]. A further annealing by AC current of the so-facetted structure restores the flatness of the nominal vicinal face. The physical interpretation is clear: when restoring AC current, the electromigration-driven instability no more works and the Joule effect restores the original stable vicinal face initially obtained by mechanical slicing. It thus confirms that the facetting we observe is a kinetic effect due to adatoms drift induced by the electrical field [22-24].

In Fig. 4 are reported, for each vicinal surface, the temporal evolution of the facetting period and of the angle $\beta$ that one of the facet of the hill-and-valley structure forms with the nominal orientation. The angle $\alpha$ of the other facet does not vary with time. ($\alpha$ and $\beta$ are shown on Fig. 5).

Three main results are worth to be underlined.

1/ The period and the orientation of the so-formed facets evolve with time. After several hundred hours of annealing, a steady-state structure formed by the so called facets $F_1$ and $F_2$ is reached. As schematically shown on Fig. 5, the $F_1$ facet is already present in the original vicinal surface so that its area simply increases with time at constant angle $\alpha$ while the $F_2$ facet slowly builts by step bunching (the angle $\beta$ and the facet area thus increase with time). In Table I are reported the vicinal Si faces we have chosen and the corresponding $F_1$ and $F_2$ facets whose crystallographic indexes have been obtained from the measurement of $\alpha$ and $\beta$.

| **Vicinal face** | **(118)** | **(223)** | **(443)** | **(510)** |
|---|---|---|---|---|
| $F_1$ (flat at the atomic scale) | (001) | (111) | (111) | (001) |
| $F_2$ (exhibit monoatomic steps) | (113) | (113) | (110) | (110) |

*Table I:* Decomposition of the vicinal faces in $F_1$ and $F_2$ facets for the stationary state.

---

[2] That means in the $[44\bar{1}]$, $[33\bar{4}]$, $[33\bar{8}]$, and $[\bar{1}50]$ directions for the $(118)$, $(223)$, $(443)$ and $(510)$ surfaces respectively.
[3] There is no facetting with DC current in the direction of descending steps.

2/ For Silicon, the $F_1$ and $F_2$ facets correspond to cusps on the free energy polar plot (see Fig. 1), that means to the two closest facets surrounding the original vicinal face (see Fig. 2).

3/ The facetting is periodic (with a weak dispersion) whatever the angle the stepped facets form with the nominal surface (see Fig. 4). This is quite consistent with the elastic origin of the periodicity (which then propagates at the sound velocity) proposed by Marchenko [12] then Alerhand et al. [13]. The mechanical equilibrium is thus insured at each step of the process. Moreover, since the facetting implies some mass transfer, a fast surface diffusion is also needed what is quite normal in the temperature range of the experiment (1373 K) where evaporation starts to take place and thus surface diffusion is highly activated. We can thus assume that the local chemical equilibrium is also achieved at each step of the process

In equilibrium conditions, the elastic origin of the periodicity has been established by [12,13] . When two facets $a$ and $b$ (characterised by their normal directions $\hat{n}_a$ and $\hat{n}_b$ and their surface stress tensors $[s]^{n_a}$ and $[s]^{n_b}$) have a common edge (characterised by its tangential unit vector $\hat{\tau}$), at the boundaries between both facets exist an elastic force $\vec{f} = \vec{s}_\tau^{n_a} - \vec{s}_\tau^{n_b}$ where $\vec{s}_\tau^{n_a} = [s]^{n_a}(\hat{\tau} \wedge \hat{n}_a)$ is the force per unit length across the common edge boundary (see Fig. 2 and Fig. 5 for t=$t_2$ and t→∞). This localised force deforms the underlying bulk and thus decreases the total energy by means of stress relaxation. The periodicity results from a competition between the positive edge energy $\rho$ and the negative stress relaxation. More precisely the wavelength reads [12,13]:

$$\lambda = \frac{2\pi c}{\sin \pi \theta} \exp\left(1 + \frac{\pi E \rho}{2 f^2 (1-\nu^2)}\right) \qquad (1)$$

where $E$ and $\nu$ are the Young modulus and the Poisson ratio of the material respectively[4], $c$ an atomic unit, $f^2 = \left(s_\tau^{n_a} - s_\tau^{n_b}\right)^2 + 4 s_\tau^{n_a} s_\tau^{n_b} \sin^2\left(\frac{\alpha+\beta}{2}\right)$ where $\alpha$ and $\beta$ are the angles the faces $a$ and $b$ form with the original orientation (see Fig. 5) and $\theta = \frac{tg\alpha}{tg\alpha + tg\beta}$ a geometrical factor. For a given facetted system the angles $\alpha$ and $\beta$ can be measured so that, for a fixed value of the edge energy $\rho$, the period $\lambda(\alpha,\beta,s_\tau^{n_a},s_\tau^{n_b})$ only depends upon the unknown values $s_\tau^{n_a}$ and $s_\tau^{n_b}$.

---
[4] Calculated for the good crystallographic orientations

Since surfaces having a symmetry axis greater than two have isotropic surface stresses[5] and owing to the chosen vicinal orientations (with common zone axis [6]), the set of the periods of the vicinal faces we consider only depends upon five unknown quantities $s_{\bar{1}10}^{001}$, $s_{\bar{1}10}^{113}$, $s_{\bar{1}10}^{111}$, $s_{\bar{1}10}^{110}$ and the anisotropic factor $\chi$ defined by $s_{\bar{1}10}^{110} = s_{001}^{110}(1+\chi)$. Thus measuring (by AFM) the period and the angles $\alpha$ and $\beta$ of a set of completely destabilised vicinal faces (labelled k) is enough to obtain a system of equations $\lambda^k(\alpha^k,\beta^k,s_{\tau_k}^{n_{a_k}},s_{\tau_k}^{n_{b_k}})$ that can be numerically solved to obtain all the unknown quantities. Moreover, the hypothesis of local equilibrium enables us to use all the other transitory (before the final steady-state state) decompositions of vicinal surfaces (whose facets orientation are simply determined by the angle they form with the mean face), to calculate the surface stress of the intermediary surface orientations. However only some of the many numerical solutions of the system have a physical meaning. In particular we consider two important physical constraints. Firstly, close to a low index orientation, step creation has an energetic cost but allows to relax the surface stress so that a face that belongs to the equilibrium shape is a minimum of surface energy but a maximum of surface stress [4]. Secondly, to the best of our knowledge, surface stresses of clean reconstructed surfaces are known to be positive [3,4,25]. We find a single numerical solution which verifies both physical constraints. It corresponds to a negative anisotropy factor ($\chi = -0.40 \pm 0.05$) quite consistent with the structure of the Si(110) surface [26]. The corresponding surface stress plot (calculated for a constant value of $\rho$ [7]) is given in Fig. 6. Notice that because of the tensorial nature of the surface stress, two branches are necessary to represent the surface stress anisotropy of the two perpendicular components $s_\tau^n$ and $s_\omega^n$ (where $\hat{\omega} = \hat{t} \wedge \hat{n}$ is the unit vector normal to the edge). However, the procedure we describe only gives access to the $s_\tau^n$ component (perpendicular to the common edge $\tau$ of the facets), that means between $[001]$ and $[110]$ directions to $s_{1\bar{1}0}^n$ and between $[110]$ and $[100]$ directions to $s_{001}^n$ (see Fig. 2). Obviously for the anisotropic (110) surface we thus have access to the two orthogonal components $s_{1\bar{1}0}^{110}$ and $s_{001}^{110}$ of the surface stress tensor $[s]^{110}$. The so-obtained surface stress values, connected by a continuous line (simple guide for the eyes) are plotted on Fig. 6.

---

[5] It is the case for the (111) surface characterised by a ternary axis, as well as for the {100} surfaces characterised by a quaternary axis. In this last case the nominal surface is known to be constituted by two type of domains of anisotropic surface stress, but for symmetrical reasons the mean stress remains isotropic.

[6] See for example $(118)$ and $(223)$ faces in Fig. 2.

[7] The introduction of reasonable edge energy anisotropy (20%) does not affect substantially the surface stress plot.

Let us underline some main points:

(1) Since in the common direction $\hat{\tau}$ (see Fig. 2) all the surfaces exhibit more or less the same local step geometry (dense row) while in the orthogonal direction $\hat{\omega}$ the local geometry of the microfacets formed by the step and the underneath terrace varies a lot, the surface stress anisotropy cannot have the same amplitude for the two branches of the surface stress plot. In other words the $s_\tau^n$ anisotropy must always be larger than the $s_\omega^n$ one, so that our procedure gives access to the polar plot of the more anisotropic branch of surface stress that means here to $s_{1\bar{1}0}^n$ anisotropy between $[110]$ and $[100]$ directions and to $s_{001}^n$ anisotropy between $[001]$ and $[110]$ directions (see Fig. 6).

(2) The surface stress anisotropy is more important than the surface energy anisotropy (compare the scales of Fig. 1 and 6). This behaviour is quite normal since it is well known that surface stress is much more sensitive to surface relaxation than surface energy [4].

(3) Open surfaces relax easier than dense ones and thus exhibit smaller surface stress. It is the case of (113) surface in comparison to (111) or (001) surface.

(4) The extrema of the surface stress plot are more abrupt for true atomic flat $\{001\}$ and $\{111\}$ faces than for stepped $\{113\}$ or $\{110\}$ faces which then only exhibit weaker extrema. As for the gamma plot for which the greater the step energy, the sharper the corresponding minimum, it is believed that the greater the stress-step[8] of the face, the sharper the maximum of the surface stress plot.

(5) Our procedure does not give access to the polar plot of the weak anisotropic branch (see point (1)). Nevertheless the surface stress anisotropy calculated for several materials [4] can be used to estimate a zone in which should appear this branch $s_\omega^n$. For this purpose we simply draw (in grey in Fig. 6) a zone in which should lie the less anisotropic branches calculated for most of these materials.

---

[8] The surface stress is the first term of the development of the surface energy with respect to strain [27,28] while the step stress is the first development of the step energy with respect to strain and the step stiffness is connected to the second derivative of the step energy with respect to orientation [28].

(6) Since the kinetics pathway towards the two closest facets $F_1$ and $F_2$ are very different, the decomposition of the original vicinal facets only give access to intermediate orientations towards faces having the slowest kinetics so that the experimental branch is incomplete in the vicinity of the $F_1$ surface. More precisely a vicinal (111) surface gives access to the stress variation near (110) or (113) orientation but not near the (111) orientation. At the same a vicinal surface of (100) only get surface stress variation near (110) or (113) but not near (100) (see Fig. 6). Thus in order to "fill the holes" of the polar plot we must use other vicinal faces. In our case we tried, without success, to use a vicinal of (113) to get the surface stress change near (111) or (001) orientations and a vicinal of (110) to get the surface stress change near (111) and (100) orientations. The reason of this failure is that it is easier to reach a stepped face by the step bunching mechanism at the basis of our procedure than to reach a flat surface for which a supplementary activation energy for step coalescence is needed. It is for this reason that using a vicinal (113) face does not enable us to explore an angular domain greater than 8° near the (111) or (100) surface[9] as yet depicted by Song et al. [29,30].

Summarizing, we have shown that the control of kinetic facetting induced by adatoms electromigration enables, for the first time, to determine the surface stress polar dependence of stable facets belonging to the equilibrium shape. This new method opens new perspectives to measure the surface stress anisotropy of all surface materials that can be destabilised by an external field and, by no means, to a better understanding of surface stress properties.

**Acknowledgements:** J.P.Astier, J.Fuhr, G.Tréglia, R.Kern

**References:**
1. Sundquist B, Acta. Metall. **12** 67-86 (1964)
2. Bermond J.M, Métois J.J, Egea X., Floret F., Surf. Sci. **330** 48-60 (1995)
3. Ibach H., Surf. Sci. Rep. **29** 193-263 (1999)
4. Müller P., Saùl A., Surf. Sci. Rep. **54** 157-258 (2004)
5. Bach C., Giesen M., Ibach H., Einstein T., Phys. Rev. Lett. **78** 4225-4228 (1997)
6. Filipetti A., Fiorentini V., Surf. Sci. **377** 112-116 (1997)
7. Olivier S., Saùl A., Tréglia G., Appl. Surf. Sci. **212/213** 866-871 (2003)
8. Olivier S., Saùl A., Tréglia G., Willaime F., to be published
9. Wynblatt P., Ku R., Surf. Sci. **65** 511-531 (1977)
10. Ibach H., Surf. Sci. **556** 71-77 (2004)
11. Berger R., Delamarche E., Lang H., Gerber C., Gimzewski J., Meyer E., Guntherodt H., Science **276** 2021-2024 (1997)

---

[9] even after 15 days of annealing at 1373 K !

12. Marchenko V., JETP **33** 381-384 (1981); JETP **54** 605-607 (1981)
13. Alerhand O., Vanderbilt D., Meade R., Joannopoulos J., Phys. Rev. Lett. **61** 1973-1976 (1988)
14. Rousset S., Repain V., Baudot G., Ellme H., Garreau Y., Etgens V., Berroir J.M., Croset B., Sotto M., Zeppenfeld P., Ferre J., Jamet J.P., Chappert C., Lecoeur J., Mat. Sci. And Eng. B **96** 169-177 (2002)
15. Wulff G., Z.Kristall. **34** 449-480 (1901)
16. C.Herring, Phys. Rev. **82** 87-93 (1951)
17. Rousset S., Pourmir F., Berroir J.M., Klein J., Lecoeur J., Hecquet P., Salanon B., Surf. Sci. **422** 33-41 (1999)
18. Johnson R. , Phys. Rev. **54** 459-467 (1938)
19. Moore A. , « Thermal facetting » in Metal surfaces: structure, energetics and kinetics (Ed. American Society for Metals, Metals Park, Ohio) 1962, p155
20. Latyshev A., Litvin L., Aseev A., Appl. Surf. Sci. **130/132** 139-145 (1998)
21. Métois J.J, Stoyanov S., Surf. Sci. **440** 407-419 (1999)
22. Yagi K., Minoda H., Degawa M., Surf. Sci. Rep. **43** 45-126 (2001)
23. Jeong H., Williams E., Surf. Sci. Rep. **34** 171-294 (1999)
24. Krug J., Dobbs H., Phys. Rev. Lett. **73** 1947 –1950 (1994)
25. D.Sander, H.Ibach, Surface free energy and surface stress, Physics of covered surfaces, in: H.Bonzel (Ed.) Landölt Bornstein New Series, **III 42,** chap. 4, (2002)
26. An T., Yoshimura M., Ono I., Ueda K., Phys. Rev. B **61** 3006-3011 (2000)
27. Shuttleworth R., Proc. Roy. Soc. London **163** 444-457 (1950)
28. Nozières P., Wolf D.E., Z. Phys. B **70** 399-407 (1988)
29. Song S., Yoon M., Mochrie S., Surf. Sci. **334** 153-169 (1995)
30. Song S., Mochrie S., Stephenson G., Phys. Rev. Lett.**74** 5240-5243(1995)

**Figures:**

**Figure 1:** 3D equilibrium shape of a silicon bulb ($[1\bar{1}0]$ zone) obtained at 1373 K) and its corresponding $\gamma$ plot (from Ref. [2])

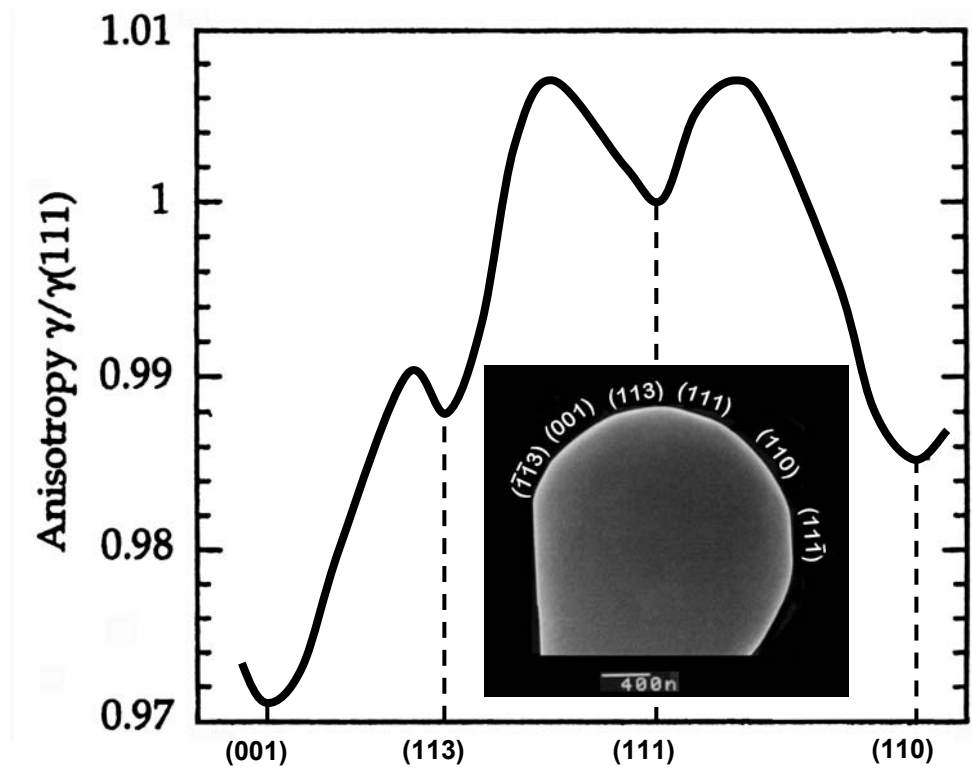

**Figure 2:** Vicinal faces under study: (**a**) main crystallographic faces and definition of the directions $\hat{n}$ (normal vector), $\hat{\tau}$ (zone axis) and $\hat{\omega} = \hat{t} \wedge \hat{n}$ of a crystalline face, (**b**) normal vectors to the vicinal faces (red) and their corresponding closest stable faces (blue).

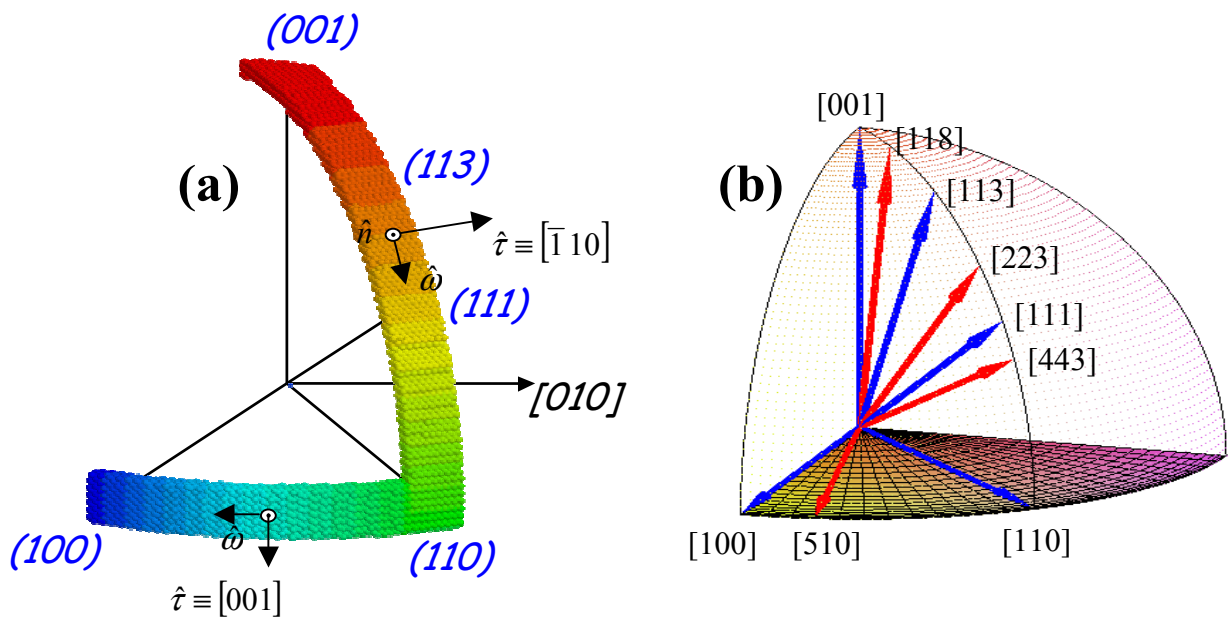

**Figure 3:** AFM image of a hill-and-valley structure obtained after annealing a (118) vicinal surface in UHV conditions at 1373 K during 150 h. The facets $F_1$ and $F_2$ are (001) and (113) respectively.

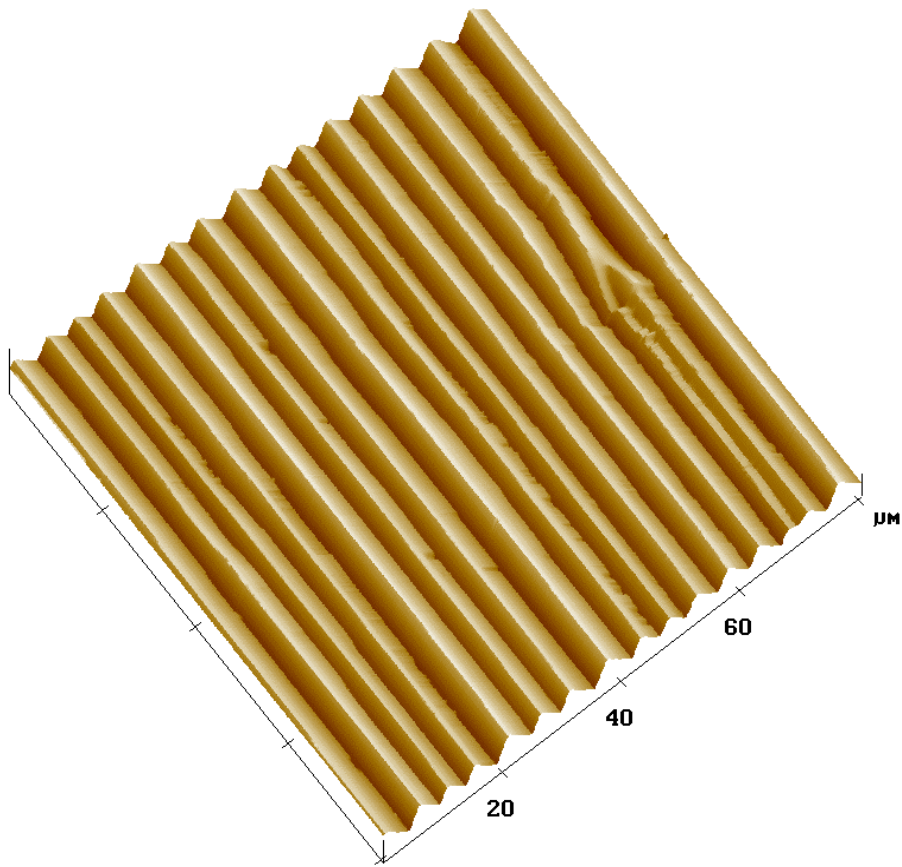

**Figure 4:** Temporal evolution of the wavelength $\lambda$ and the crystallographic angle $\beta$ for all the studied vicinal faces (the $\alpha$ angle which does not evolve with time is not reported). Line are only guides for the eyes and the error-bars correspond to experimental reproducibility.

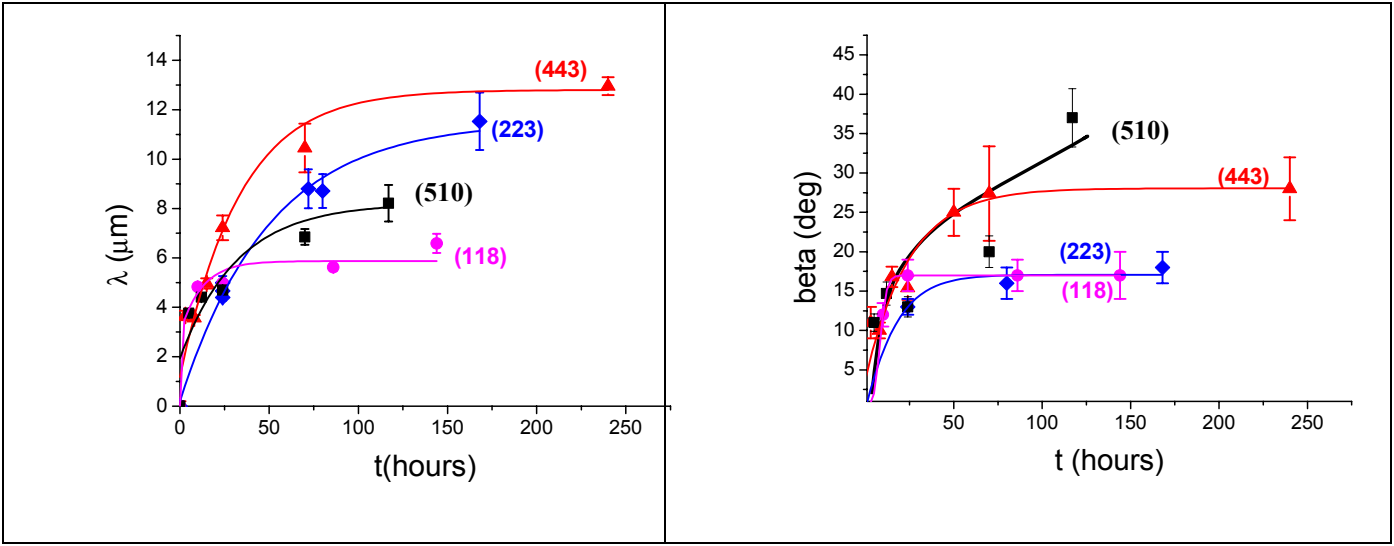

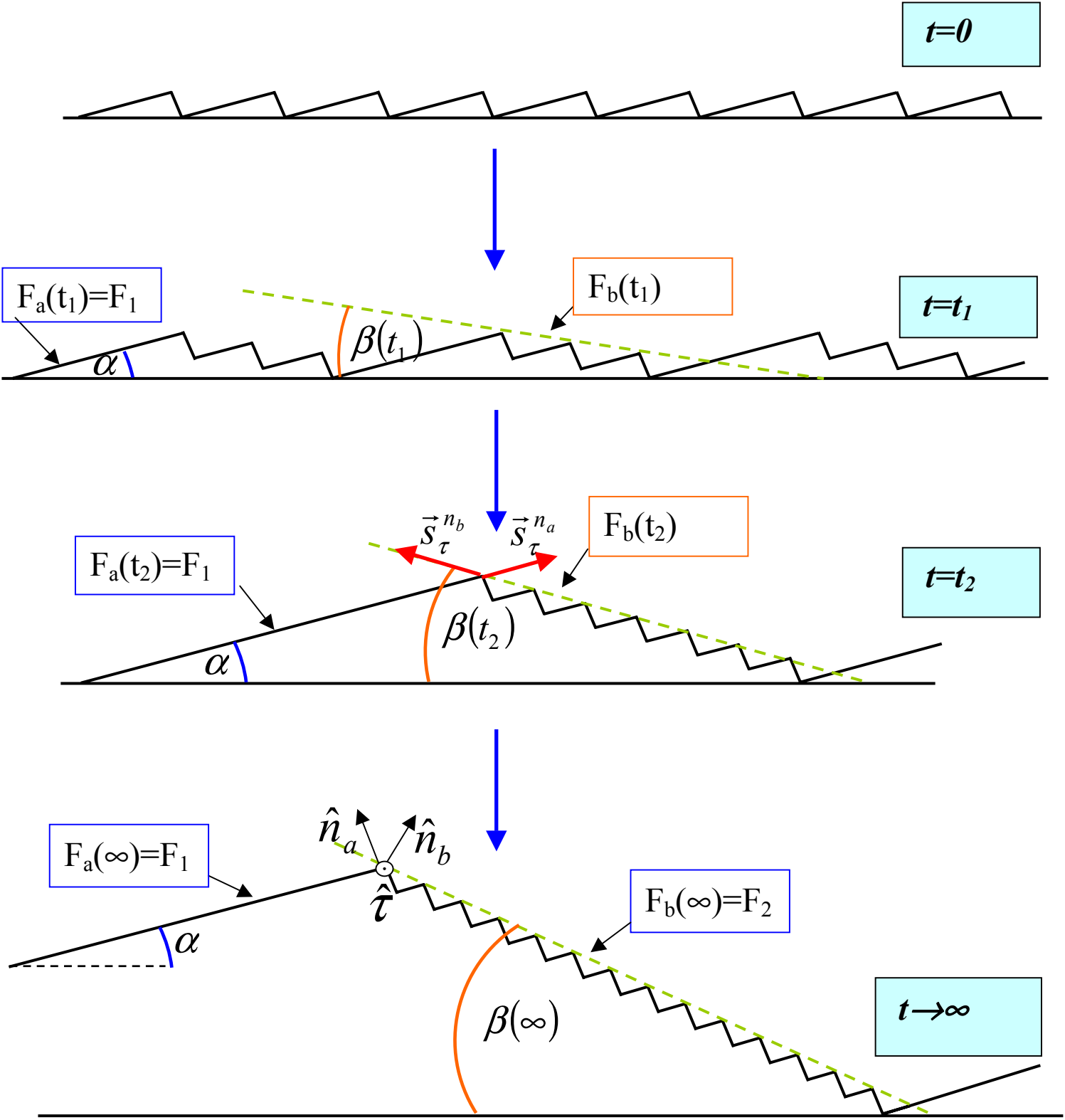

**Figure 5:** Sketch of the mechanism of kinetics facetting (t is the time)

**Figure 6** : Surface stress polar dependence (in units of $\sqrt{\dfrac{E\rho}{1-\nu^2}}$ calculated for the (100) face).

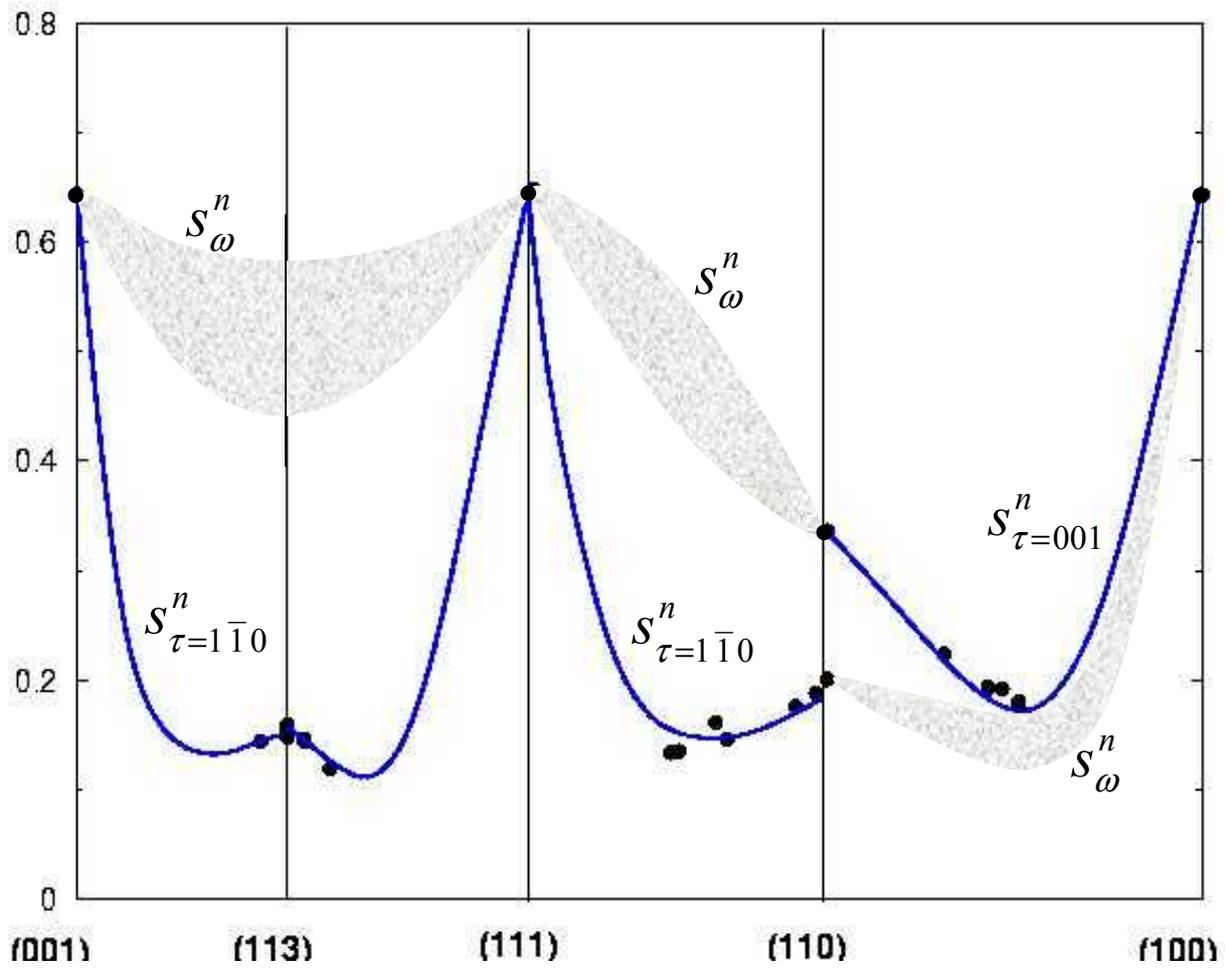